# Confinement of fractional quantum Hall states in narrow conducting channels

R.L. Willett, M.J. Manfra, L.N. Pfeiffer, K.W. West
Bell Laboratories, Alcatel-Lucent

ABSTRACT:

Confinement of small-gapped fractional quantum Hall states facilitates quasiparticle manipulation and is an important step towards quasiparticle interference measurements. Demonstrated here is conduction through top gate defined, narrow channels in high density, ultra-high mobility heterostructures. Transport evidence for the persistence of a correlated state at filling fraction 5/3 is shown in channels of 2μm length but gated to near 0.3μm in width. The methods employed to achieve this confinement hold promise for interference devices proposed for studying potential non-Abelian statistics at filling fraction 5/2.

INTRODUCTION:

Confinement of a correlated electron state and control of its charge excitations provides an important experimental avenue for examining quasiparticle properties. Conduction through quantum point contacts [1] (qpcs), narrow channels [2], and hybrids of qpcs and channels [3] has provided data potentially describing quasiparticle charge through noise and interference effects. These experiments have studied lowest Landau level fractional quantum Hall states with large energy gaps at primary filling factors 1/3 and 2/5. Further efforts in elaborate devices again combining conducting channels and throttling qpcs[4] have been successful at examining interference phenomena for integer quantum Hall states (IQHE), but have not demonstrated such effects for fractional quantum Hall effect (FQHE) states. A particular challenge is to produce high aspect ratios devices, such as long narrow channels, that provide a small lateral dimension to access length scales that may be associated with quasiparticle coherence, yet long enough to present experimentally accessible charge flow paths

It is an important goal to produce such confinement devices, such as channels, quantum point contacts, and devices combining these components with the ability to sustain *small* gapped states as a tool to understand these states and their excitations. As an example, confinement of the quantum Hall state at filling factor 5/2 and interference of its quasiparticles in a channel/qpc device has been proposed as a means of determining whether the state displays non-Abelian statistics [5]. In a similar device quantum computation manipulations have been proposed [6]. Small gapped exotic FQHE states have been observed only in ultra-high mobility heterostructures following minimal processing and efforts to preserve large (mm scale) areas over which transport is performed [7]. With no previous demonstration of mesoscopic confinement of FQHE states other than those with large excitation gaps, small gapped state confinement is complicated by several deleterious device processing effects: metals on the sample surface, gate defined non-smooth edges, and overall processing procedures that alter the sample surfaces.

Presented here are the results of experiments accomplishing confinement of fractional quantum Hall states in narrow, tunable channels with high aspect ratios. Using high mobility, high density heterostructures, top gate devices ranging from large (10μm x 20μm, lithographically defined) to small (1μm x 2μm) produce narrow conducting channels able to support FQHE states. Channel properties were first assessed by examining the B=0 diffusive transport for the different channel dimensions, with each device tunable in width via applied gate voltage. With focus on the 1μm x 2 μm channel, FQHE properties of zero longitudinal resistance and quantized Hall resistance are apparent for filling factor 5/3 even for tuning of the channel to a width of near 0.3μm, while maintaining the 2 μm length. In addition, transport data of the N=1 (filling fraction ν=2 to 4) Landau level including the filling factors 5/2, 7/3, and 8/3 are also recorded for the smallest channel, with width tuned to near 0.3μm. This method of operation using high density, high mobility, top-gated structures presents a viable path to confinement of fragile correlated electron states.

## METHODS:

The heterostructures used in this study are ultra-high mobility (> 15 x $10^6$ $cm^2$/V-sec) quantum wells constructed with densities exceeding 4 x $10^{11}$/$cm^2$. The use of higher density is a departure from previous confinement experiments [1-4]. The development of ultra-high mobility, high density two-dimensional electron gases will be detailed in an upcoming publication. The quantum well is roughly 200nm below the surface of the heterostructures. Following photolithographic patterning of mesas and evaporation and diffusion of Ni/Au/Ge contacts, the top gates that define the channels are patterned using e-beam lithography (see Figure 1). The metallic top gates are aluminum, deposited first slowly to achieve an oxide layer, then rapidly to produce the metallic layer. A range of channel dimensions was initially tested, all with aspect ratios of 2:1. The dimensions were 10μm x 20 μm, 4μm x 8 μm, 2μm x 4 μm, and 1μm x 2 μm. The focus of the study is on the narrowest 1μm x 2 μm channels. In operation, the top gates were typically able to support ~ -6 volt bias without demonstrable leakage to the 2D layer. Standard low noise transport measurements were performed using lock-in techniques in both He3 and dilution refrigeration systems.

## RESULTS:

The first focus of measurements was on coarse operation of the gates and determining the channel widths achievable with gate bias. A metric of this width is the B=0 resistance change that occurs for application of gate bias. With near -1V applied to the top gate structures depletion of the underlying 2D electron gas is achieved, leaving a channel roughly the lithographically defined dimension. For the channels 4μm x 8μm, 2μm x 4μm, and 1μm x 2μm, further negative bias of the gate produces a change in width sufficient to increase the B=0 resistance due to the diffuse boundary scattering at the channel edges [8]. As anticipated, this biasing past depletion has the largest effect for the narrowest lithographically defined channel (1μm x 2 μm), and it is these results that are described here. Figure 1 shows longitudinal magneto-transport for such a device under different gate biases, demonstrating the striking suppression of the B=0 resistance upon application of a magnetic field; this can be understood to occur due to the reduction of diffuse boundary backscattering where the B-field initially at low values promotes boundary scattering, then at higher values induces orbits traversing the channel edge, thus reducing the channel resistance. For sufficiently narrow channels only a monotonic decrease in resistance should be observed [8]. The data of Figure 1 show a substantially increasing B=0 resistance over the range of biases applied. The peak values of the resistance at B=0 are plotted in Figure 1c as a function of applied gate voltage, demonstrating the depletion and progressive narrowing of the channel for larger negative applied bias. For such a diffuse boundary scattering channel it is appropriate [8] to assign the resistance increase to a reduction in width using $R = R_0(1+a/W)$, where W is the functional or biased width of the channel and a is determined at onset of full depletion just under the gate. The enhancement of the resistance by the decreasing width is apparent in the data, and using the nominal lithographic width at depletion of 1μm, the derived width at higher gate biases using this relationship is marked by sample values in the figure. As such, the gate bias producing the magneto-transport trace of Figure 1b showing properties to filling factor 2 corresponds to a channel widths decreasing with bias down to below 0.25 μm but retaining the length of 2μm.

Following determination of the channel width as a function of gate bias, the measurement goal is then to assess what states can exist within the constrictions upon gate bias. While the data of Figure 1b show a four terminal (T=0.3K) magneto transport trace through the constriction with a range of IQHE and potentially FQHE features, the bulk contribution to this measurement may dominate the results. The principal route to assess the constriction and overcome the bulk contribution is the following: examination of the Hall trace through the constriction will show proper quantization reflecting the number of transmitted channels if the bulk longitudinal resistivity is zero for those states and the state is preserved within the channel. This condition would prevail for robust FQHE states. In order to assess transport through the channel, the quantities $R_L$ and $R_D$ are measured [8]. $R_L$ is a longitudinal resistivity measurement which includes traversal of the channel by using voltage probes on the same side of the sample but separated by the channel. $R_D$ corresponds to the topologically equivalent Hall voltage measurement but also across the channel, thus necessarily including a contribution from the channel's longitudinal resistance. These measurement configurations are shown schematically in Figure 1a. Within the Landaur-Buttiker formalism [8] it is expected that the longitudinal resistance should follow $R_L = h/e^2(1/N_{min} - 1/N_{wide})$, where $N_{min}$ and $N_{wide}$ represent the number of quantum Hall edge states in the channel and in the bulk, respectively, with the bulk Hall resistance $R_H = h/e^2(1/N_{wide})$, and $R_D = R_L + R_H$ for a B-field orientation used here.

Magneto-transport through the smallest channel (1μm x 2 μm) is shown in Figure 2 at dilution refrigerator temperatures and compared to measurements through the bulk. Note that for all biasing conditions, quantization is maintained at filling factors 3 and 2. Our principal finding for the 1μm x 2 μm device is that upon negative biasing of the gate past full depletion, the FQHE state at filling factor 5/3 persists. $R_L$ demonstrates a minimum at or near zero within the measurement capabilities, showing persistence of the states over the bulk and through the channel. $R_D$ also shows quantization to the appropriate value upon biasing, consistent with persistence of the state again across the channel. Both $R_D$ and $R_L$ demonstrate these features through a large range of gate bias, from depletion to -4.5V. At the largest negative bias this data provides evidence that the 5/3 state is robust within the channel at a gate voltage corresponding to a channel width of roughly 0.3μm, with the channel length 2μm or somewhat longer due to the lateral depletion at the ends of the channel. We measured the activation energy for the bulk 5/3 state to be approximately 1.8K. Given that the 5/3 state is supported within the channel, this activation energy estimates the gap energy scale that can be confined in a channel of these extreme dimensions.

Attention is now turned to the fragile states in the N=1 Landau level, those at filling factors 5/2, 7/3, and 8/3 for the 1μm x 2 μm channel. The measurements of $R_D$ and $R_L$ over these filling factors are also shown in Figure 2, again compared to the bulk measurements. Note in the bulk measurements prominent minima in $R_{xx}$ and developed plateaus in $R_{xy}$ are apparent for the fractional states at 5/2, and 7/3. As gate bias is increased throughout the panels of Figure 2, the longitudinal features in $R_L$ at these filling factors persist, but on an increasing background: the corresponding measurement of $R_D$ shows the developing plateaus at these filling factors displaced from their proper quantization levels. Note that this displacement corresponds roughly to the background increase in $R_L$. This residual displacement of the Hall trace is presumably due to contribution of the longitudinal resistance to the $R_D$ within the channel. Whether the channel would support these fragile states at lower temperatures is an open question: at sufficiently low temperatures where $R_{xx}$ is near zero for these states in the bulk, $R_D$ in the channel could show proper quantization given a near zero longitudinal contribution. In examining the Hall resistance through the channel for a specific B-field direction, the bulk Hall resistance $R_H$ is indeed recovered as it should be when the difference of the measured resistances is used: $R_H = R_D - R_L$.

CONCLUSION:

The method of using high mobility, high density heterostructures with specific surface gates to define small channels demonstrates successful confinement of fractional quantum Hall states, potentially into the N=1 Landau level. The persistence of the FQHE states in a channel with high operating aspect ratio of ~0.3μm x 2μm suggests that the integrity of the edge states can be maintained with severe lateral confinement over relatively large length-scales for less robust fractional states. This channel length dimension of 2μm is consistent with a device dimension where further elaboration of parts to facilitate interference phenomena can be overlain through precise but existent alignment methods. Such elaborations and lower temperature examinations are presently underway.

REFERENCES:

[1] R. de-Picciotto, M. Reznikov, M. Heiblum, V. Umansky, G. Bunin, D. Mahalu, Nature **389**, 162-164 (1997).

[2] J.A. Simmons, H.P. Wei, L.W. Engel, D.C. Tsui, M. Shayegan, Phys. Rev. Lett **63**, 1731 (1989).

[3] F.E. Camino, W. Zhou, V.J. Goldman, Phys. Rev. Lett. **95** (24), 246802 (2005).

[4] Y. Ji, Y. Chung, D. Sprinzak, M. Heiblum, D. Mahalu, H. Shtrikman, Nature **422**, 415 (2003).

[5] A. Stern, B.I. Halperin, Phys. Rev. Lett. **96**, 016802 (2006).

[6] S. Das Sarma, M. Freedman, C. Nayak, Phys. Rev. Lett. **94**, 166802 (2005).

[7] R.L. Willett, J.P. Eisenstein, H.L. Stormer, D.C. Tsui, A.C. Gossard, J.H. English, Phys. Rev. Lett. **59**, 1776 (1987). See also W.Pan, J.S. Xia, V. Shvarts, D.E. Adams, H.L. Stormer, D.C. Tsui, L.N. Pfeiffer, K.W. Baldwin, K.W. West, Phys. Rev. Lett., **83** (17), 3530 (1999) .
[8] C.W.J. Beenaker and H.van Houten, Solid State Physics **44**, 1-128 (1991).

FIGURE CAPTIONS:

Figure 1. top gate channel structures: a) lead frame configuration around the top-gate channel, left, also showing contact arrangements for $R_L$ and $R_D$ measurements for current driven from contact 8 to 4; scanning electron micrograph of a 4μm x 8μm channel. b) d.c. magneto-transport ($R_L$) through a 1μm x 2μm channel showing resistance increase at B=0 due to diffusive boundary scattering as the channel is narrowed by top-gate biasing, suppressed by magnetic field, with bias ranging from 0 to -5.3V. The temperature is 280mK. The large range B-field trace extends to opposite field directions, the remaining traces are reflected about B=0 for demonstration purposes. c) B=0 resistance values for different top gate biases; several channel widths are marked corresponding to a fit consistent with diffusive boundary scattering in a channel [8]: R(B=0) = 200Ω + 400Ω[0.5μm/(1μm-(V-1)0.18μm/V)].

Figure 2. evidence for the FQHE state at filling factor 5/3 propagating through 1μm x 2μm channel: (a) Hall and longitudinal resistances for conduction in the bulk; (b) $R_L$ and $R_D$ measurements at bias = -1.2V, corresponding to depletion under the gate, and a width of 1μm (c) $R_L$ and $R_D$ measurements at bias = -3.5V, corresponding to a width of ~0.5μm. (d) $R_L$ and $R_D$ measurements at bias = -4.5V, corresponding to a width of ~0.3μm. Note the preservation of quantization at 5/3 in $R_D$. Temperature is 50mK for all traces.

Figure 1.

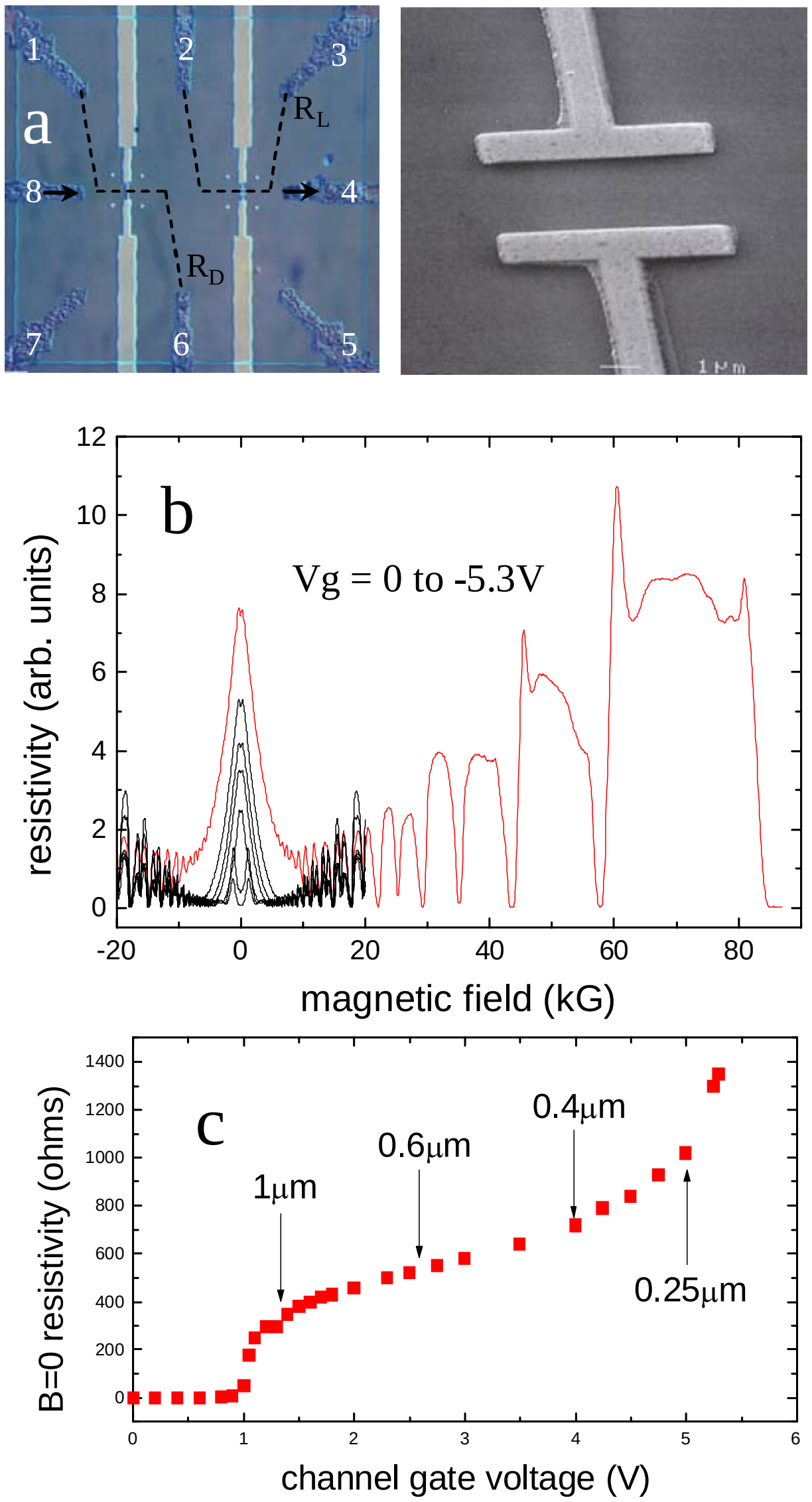

Figure 2.

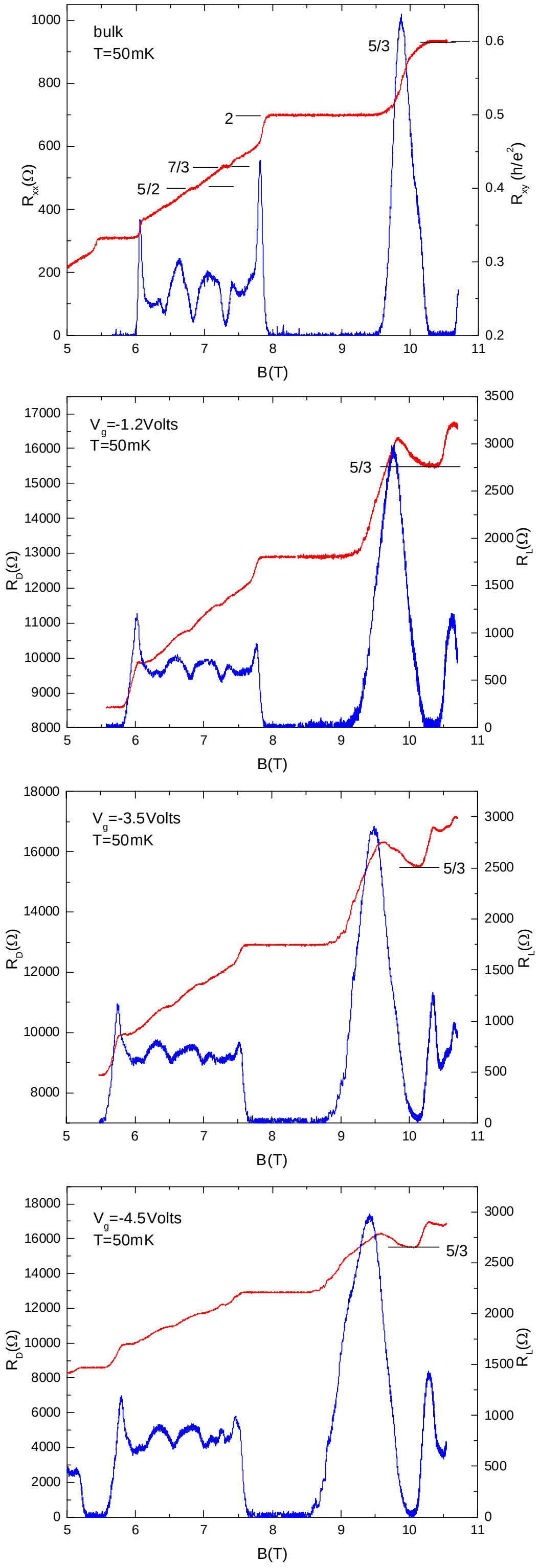